\RequirePackage{fix-cm}
\documentclass[twocolumn,epjc3,comma,sort&compress,natbib]{svjour3}
\bibpunct{[}{]}{,}{n}{}{,} 

\journalname{Eur. Phys. J. A}

\usepackage{latexsym}
\usepackage{amsmath}
\usepackage{amssymb}
\usepackage{amsfonts}

\usepackage[mathscr,scaled=1.15]{urwchancal}
\DeclareFontFamily{OT1}{pzc}{}
\DeclareFontShape{OT1}{pzc}{m}{it}%
{<-> s * [1.15] pzcmi7t}{}
\DeclareMathAlphabet{\mathpzc}{OT1}{pzc}{m}{it}

\usepackage{color}

\usepackage{supertabular}
\usepackage{placeins}
\usepackage{epsfig}
\usepackage{graphicx}

\definecolor{purple}{rgb}{0.5,0,0.5}
\definecolor{blue}{rgb}{0.0,0,0.9}
\definecolor{prdblue}{rgb}{0.133,0.118,0.498}
\usepackage[colorlinks=true, pdfstartview=FitV, linkcolor=prdblue, citecolor= prdblue, urlcolor=prdblue]{hyperref}

\begin{document}

\title{$\,$\\[-7ex]\hspace*{\fill}{\normalsize{\sf\emph{Preprint no}. NJU-INP 038/21}}\\[1ex]
Distribution amplitudes of light diquarks
}

\author{Ya Lu\thanksref{NJU,INP}
       \and
       Daniele Binosi\thanksref{ECT}
        \and
       Minghui Ding\thanksref{ECT} 
       \and \\
       Craig D. Roberts\thanksref{NJU,INP}
       \and
       Hui-Yu Xing\thanksref{NJU,INP}
       \and
       Chang Xu\thanksref{NJU,INP}
}


\authorrunning{Ya Lu \emph{et al}.} 

\institute{School of Physics, Nanjing University, Nanjing, Jiangsu 210093, China \label{NJU}
           \and
           Institute for Nonperturbative Physics, Nanjing University, Nanjing, Jiangsu 210093, China \label{INP}
           \and
           European Centre for Theoretical Studies in Nuclear Physics and Related Areas, \\ \mbox{$\;\;$}Villa Tambosi, Strada delle Tabarelle 286, I-38123 Villazzano (TN), Italy \label{ECT}\\[1ex]
Email:
\href{mailto:luya@nju.edu.cn}{luya@nju.edu.cn} (Y. Lu);
\href{mailto:binosi@ectstar.eu}{binosi@ectstar.eu} (D. Binosi);
\href{mailto:binosi@ectstar.eu}{mding@ectstar.eu} (M. Ding);\\
\hspace*{3.1em}\href{mailto:cdroberts@nju.edu.cn}{cdroberts@nju.edu.cn} (C. D. Roberts);
\href{mailto:hyxing@smail.nju.edu.cn}{hyxing@smail.nju.edu.cn} (H.-Y. Xing);
\href{mailto:cxu@nju.edu.cn}{cxu@nju.edu.cn} (C. Xu)
            }

\date{2021 March 04}

\maketitle

\begin{abstract}
Accumulating evidence indicates that soft quark+quark (diquark) correlations play an important role in the structure and interactions of hadrons constituted from three or more valence-quarks; so, it is worth developing insights into diquark structure.  Using a leading-order truncation of those equations needed to solve continuum two-valence-body bound-state problems, the leading-twist two-parton distribution amplitudes (DAs) of light-quark scalar and pseudovector diquarks are calculated.  The diquark DAs are narrower and taller than the asymptotic profile that characterises mesons.  Consequently, the valence quasiparticles in a diquark are less likely to carry a large light-front fraction of the system's total momentum than those in a meson.  These features may both influence the form of baryon DAs and be transmitted to diquark distribution functions (DFs), in which case their impact will be felt, \emph{e.g}.\ in the proton's $u$ and $d$ valence-quark DFs.
\end{abstract}



\noindent\textbf{1.$\;$Introduction}.\,---\,%
Modern experimental facilities and wide ranging theoretical studies have provided strong indications that soft\footnote{``soft'' means the correlations have electromagnetic sizes typical of mesons; hence, they are not pointlike diquarks.}  quark+quark (diquark) correlations play a prominent role in hadron structure \cite{Barabanov:2020jvn}.  For instance, theory suggests that diquark correlations form as a sure consequence of dynamical chiral symmetry breaking (DCSB); namely, as a corollary of emergent hadronic mass (EHM), itself responsible for almost all visible mass in the Universe \cite{Roberts:2021xnz}.
Moreover, phenomenology points to a possible role for diquarks in explaining the appearance of tetra- and penta-quark hadrons, and experiments have revealed signals for diquark correlations in, amongst other things, the proton's flavour-separated electromagnetic form factors \cite{Barabanov:2020jvn}.

Regarding diquark-based descriptions of the proton, whose valence-quark content is two $u$ quarks and one $d$ quark, an isoscalar-scalar diquark, $[ud]$, and two isovector-pseu\-do\-vector diquarks, $\{ud\}$, $\{uu\}$, are \emph{necessary} and \emph{sufficient} to explain its structure and interactions.  Other diquark correlations are possible, \emph{e.g}.\ isoscalar-pseudo\-scalar; but analyses show all such additional correlations to be irrelevant so far as proton properties are concerned \cite{Eichmann:2016yit, Qin:2020rad, Yin:2019bxe, Yin:2021uom}.  The $[ud]$ is dominant, providing roughly 60\% of the proton's normalisation \cite{Barabanov:2020jvn}; but $\{ud\}$, $\{uu\}$ are crucial because, \emph{inter alia}, their presence enables the $d$ valence-quark to participate in hard interactions, as illustrated elsewhere \cite[Sec.\,3.6]{Chen:2020ijn}.  Theories of the proton that omit pseudovector correlations are in conflict with quantum chromodynamics (QCD); indeed, they are kindred to models of the meson spectrum that contain the pion but exclude the $\rho$-meson.

This last remark is supported by the observation that at leading-order (rainbow-ladder, RL) in a widely-used symmetry-preserving truncation scheme for the continuum bound-state problem \cite{Munczek:1994zz, Bender:1996bb}, clear and marked analogies can be drawn between $J^P$ mesons and $J^{-P}$ diquarks.  Thus, for many reasons, it is useful to consider pairings, \emph{e.g}.\ pseudoscalar mesons $\leftrightarrow$ scalar diquarks, and vector mesons $\leftrightarrow$ pseudovector diquarks.  Regarding diquark mass scales, this is highlighted in Ref.\,\cite[Fig.\,3]{Yin:2021uom}.

The presence of diquarks within hadrons is manifested in many ways \cite{Barabanov:2020jvn}; but basic signals, characteristics and insights can be drawn from analyses of hadron wave functions.  The proton is the simplest system in which diquarks appear; so its wave function is a sensible place to begin.

In quantum field theory, proton structure is described by a Poincar\'e covariant Faddeev amplitude, $\Psi$.  As explained and illustrated elsewhere \cite{Eichmann:2016yit, Qin:2020rad}, in concert with realistic interaction currents \cite{Oettel:1999gc, Chen:2020wuq, Chen:2020:progress}, knowledge of $\Psi$ enables predictions to be made for measurable proton properties.  However, $\Psi$ does not have the probability interpretation of a Schr\"odinger wave function in quantum mechanics.

A hadron's light-front wave function has a probability interpretation \cite{Brodsky:1997de, Hiller:2016itl, Mannheim:2020rod}.  This quantity can be obtained from $\Psi$ by light-front projection \cite{Heinzl:2000ht}, as illustrated by the calculation of the proton's leading-twist dressed-valence-quark distribution amplitude (DA) in Ref.\,\cite{Mezrag:2017znp}.  As evident therein, the pointwise behaviour of the proton's DA depends upon the functional forms of the analogous DAs of the active diquark correlations within the proton.  These functions depend on $x$, the light-front longitudinal momentum-fraction of a quark within the diquark.  (The other quark carries $\bar x = 1-x$.)  Hitherto, these quantities have been modelled, but not calculated.  Herein, we remedy that by presenting the first predictions for the $x$-dependence of diquark DAs.

\smallskip

\noindent\textbf{2.$\;$Insights from a contact interaction}.\,---\,%
Before completing a numerical study of the meson and diquark bound-state problems using realistic kernels, it is worth using a related, algebraic framework in order to learn what might reasonably be expected.  Such is provided by a symmetry-preserving formulation of a vector$\times$vector contact interaction (CI) \cite{Roberts:2011wy}.  Working in the chiral limit, so that the current-masses of the $u$ and $d$ valence quarks are zero, and solving the gap equation using an interaction strength that produces a reasonable value of the pion's leptonic decay constant ($f_\pi=0.10\,$GeV), one obtains a $u=d$ dressed-quark mass $M=0.36\,$GeV -- a scale typical of QCD \cite{Binosi:2016wcx, Cyrol:2017ewj, Aguilar:2018epe, Serna:2018dwk, Oliveira:2020yac}.  The CI dressed-quark propagator is $S(k) = 1/[i\gamma\cdot k + M]$.

Using the CI, the pion and $0^+_{[ud]}$-diquark are described by correlation amplitudes with the following forms, where $P$ is the total momentum of the system:
\begin{subequations}
\label{BSamplitudes}
\begin{align}
\Gamma_\pi(P) & = \gamma_5 \left[ i E_\pi(P) + \frac{1}{2M} \gamma\cdot P F_\pi(P)\right]\,,
\label{piBSA}\\
\Gamma_{0^+}(P)C^\dagger & = \gamma_5 \left[ i E_{0^+}(P) + \frac{1}{2M} \gamma\cdot P F_{0^+}(P)\right]\,.
\label{qqBSA}
\end{align}
\end{subequations}
Here, $C$ is the charge conjugation matrix.  Since the contact interaction is momentum-independent, the amplitudes do not carry any dependence on the relative momentum between the valence constituents.

The amplitudes in Eqs.\,\eqref{BSamplitudes} satisfy algebraic Bethe-Sal\-pe\-ter equations \cite[Eq.\,(12), Eq.\,(25)]{Yin:2019bxe}, which differ only by a relative multiplicative factor of one-half in the diquark equation.  In the chiral limit, these equations yield the following masses and canonically normalised Bethe-Salpeter amplitudes (masses in GeV):
\begin{equation}
\label{BSsolutions}
\begin{array}{c|c|c||c|c|c}
m_\pi & E_\pi & F_\pi & m_{0_{[ud]}^+} & E_{0^+} & F_{0^+} \\\hline
0 & 3.56 & 0.46 & 0.77 & 2.72 & 0.30
\end{array}\,.
\end{equation}
Despite the large difference in mass, arising because the chiral-limit pion is a Nambu-Goldstone mode, the $\pi$ and $0_{[ud]}^+$ Bethe-Salpeter amplitudes are semiquantitatively similar, possessing the same sign and ordering, but with the $0_{[ud]}^+$ amplitude being roughly 30\% smaller numerically.

With dressed-quark propagators and Bethe-Salpeter amplitudes in hand, the associated leading-twist DAs can be computed via light-front projection.  For the pion \cite{Chang:2013pq}:
\begin{subequations}
\label{varphiresult}
\begin{align}
f_\pi \,\varphi_{\pi}(x;\zeta)& = N_c {\rm tr}_{\rm D}Z_2(\zeta,\Lambda) \nonumber\\
& \times \int_{dk}^\Lambda \delta_n^x(k_\eta)\gamma_5 \gamma\cdot n \chi_\pi(k_{\eta\bar\eta},P ;\zeta)\,,\\
\chi_\pi(k_{\eta\bar\eta},P;\zeta) & =
S(k_\eta;\zeta) \Gamma_\pi(k_{\eta\bar\eta},P;\zeta) S(k_{\bar \eta};\zeta)\,.
\label{defchi}
\end{align}
\end{subequations}
Here we write general formulae, valid when the Bethe-Salpeter amplitudes are momentum-dependent, and
$N_c$ $=$ $3$; the trace is over spinor indices;
$\int_{dk}^\Lambda$ is a symmetry-preserving regularisation of the four-dimen\-sio\-nal integral, with $\Lambda$ the regularisation scale;
$Z_2$ is the quark wave function renormalisation constant, with $\zeta$ the renormalisation scale;
$\delta_n^x(k_\eta) = \delta(n\cdot k_\eta - x n\cdot P)$, $n$ is a light-like four-vector, $n^2=0$, with $n\cdot P = -m_\pi$ in the meson rest frame;
$\bar n$ is a conjugate light-like four-vector, $\bar n^2 = 0$, $n\cdot\bar n = -1$;
$k_{\eta\bar\eta}=[k_\eta + k_{\bar \eta}]/2$, $k_\eta = k+\eta P$, $k_{\bar\eta}=k-(1-\eta)P$;
and $f_\pi$ is the pion's leptonic decay constant, so
\begin{align}
\label{unitnormpi}
\int_0^1 dx\, \varphi_{\pi}(x;\zeta) = 1\,.
\end{align}
For future reference, a dependence on the resolving scale, $\zeta$, is also indicated explicitly here.  Typically, however, it will be suppressed.

For the $0^+_{[ud]}$ system, the analogous expression is
\begin{subequations}
\label{varphiud}
\begin{align}
f_{0^+} \,\varphi_{0^+}(x;\zeta)& = N_c^{\bar 3} {\rm tr}_{\rm D}Z_2(\zeta,\Lambda) \nonumber\\
& \times \int_{dk}^\Lambda \delta_n^x(k_\eta)\gamma_5 \gamma\cdot n \chi_{0^+}^{C^\dagger}(k_{\eta\bar\eta},P ;\zeta)\,,\\
\chi_{0^+}^{C^\dagger}(k_{\eta\bar\eta},P) & =
S(k_\eta) \Gamma_{0^+}(k_{\eta\bar\eta},P)C^\dagger S(k_{\bar \eta})\,.
\end{align}
\end{subequations}
In this case, the trace over colour gives $N_c^{\bar 3}=2$ because the diquarks considered herein are colour-antitriplet correlations.  Again, $f_{0^+}$ is defined such that
\begin{align}
\label{unitnormud}
\int_0^1 dx\, \varphi_{0^+}(x) = 1\,.
\end{align}

Using Eqs.\,\eqref{piBSA}, \eqref{BSsolutions}, \eqref{varphiresult}, Ref.\,\cite[Sec.\,III\,C]{Roberts:2010rn} computed the contact-interaction pion DA:\footnote{In such contact interaction studies, $\Lambda = \zeta$ plays a dynamical role, defining the range of interactions, and $Z_2\to 1$.}
\begin{equation}
\label{piDAans}
\varphi_\pi(x) = \Theta(x) \Theta(\bar x)\,,
\end{equation}
where $\Theta(x)$ is the Heaviside function.  The same result is obtained in symmetry preserving treatments of the Nambu--Jona-Lasinio model, \emph{e.g}.\ Ref.\,\cite{RuizArriola:2002bp}.  This distribution produces the following Mellin moments:
\begin{subequations}
\begin{align}
\langle (x-\bar x)^{2 m} \rangle_\pi & = \int_0^1 \! dx\, (x-\bar x)^{2m} \varphi_\pi(x) \\
& = \frac{1}{1+2 m}\,,
\end{align}
\end{subequations}
$m\in {\mathbb Z}^\geq$.  All odd moments vanish.

Using Eqs.\,\eqref{qqBSA}, \eqref{BSsolutions}, \eqref{varphiud}, one readily obtains
\begin{equation}
\label{qqmoments}
\begin{array}{c|c|c}
\langle (x-\bar x)^{0} \rangle_{0^+} & \langle (x-\bar x)^{2} \rangle_{0^+} & \langle (x-\bar x)^{4} \rangle_{0^+}\\\hline
1 & 0.27 & 0.15
\end{array}\,.
\end{equation}
Noting the form of $\varphi_\pi(x)$ in Eq.\,\eqref{piDAans} and capitalising on the fact that Gegenbauer polynomials of degree ``$\tfrac{1}{2}$'' form a complete orthonormalisable set with respect to the weight function ${\mathpzc w}(x)=\,$constant, one can use the moments in Eq.\,\eqref{qqmoments} to construct an approximation to the $0^+_{[ud]}$ DA:
\begin{equation}
\label{qqDAans}
\varphi_{0^+}(x) = 1 - \tfrac{19}{40} C_2^{1/2}(2 x-1) + \tfrac{9}{100} C_4^{1/2}(2 x-1)\,.
\end{equation}

\begin{figure}[t]
\includegraphics[clip, width=0.42\textwidth]{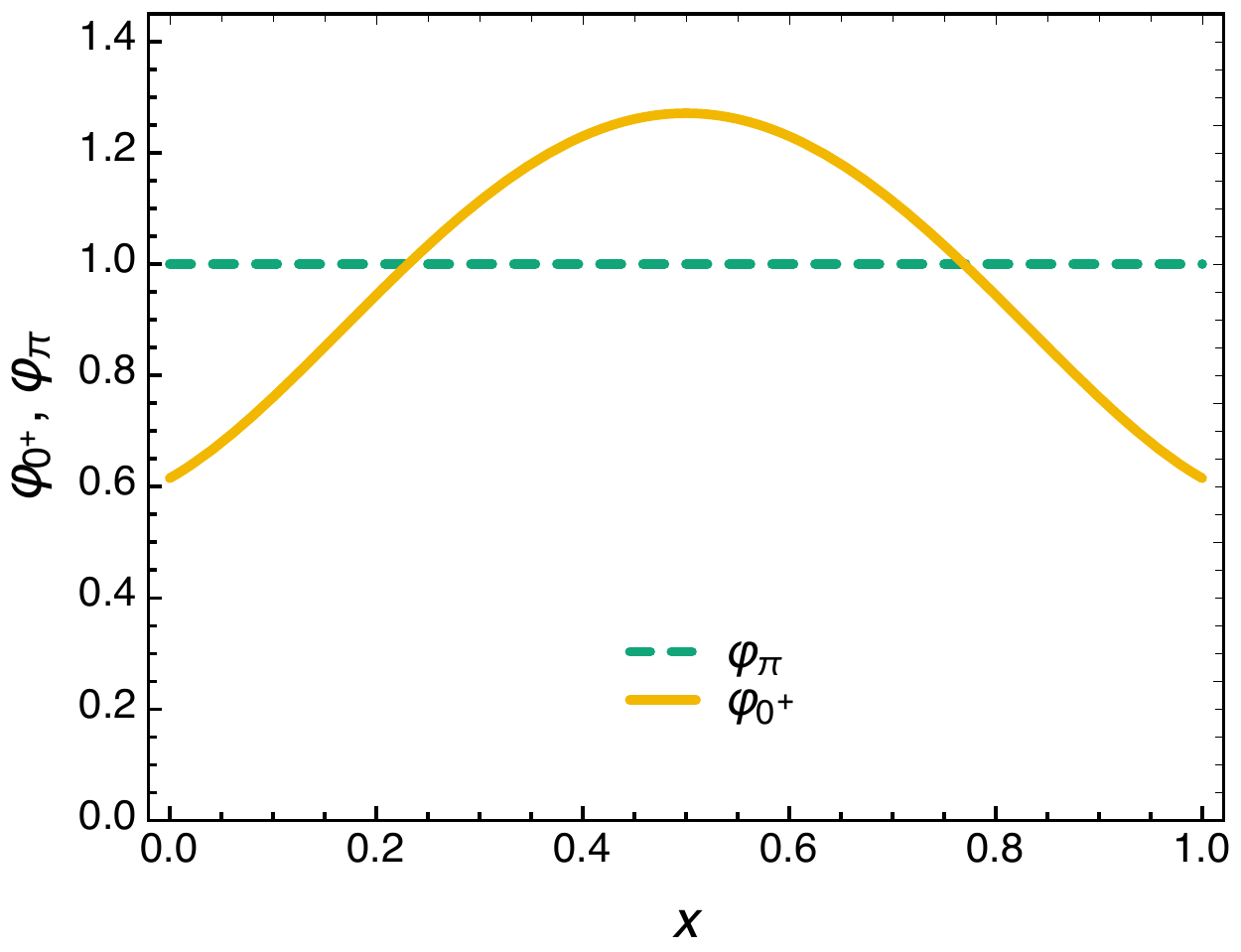}
\caption{\label{FigCIDAs}
Two-particle DAs for the pion and scalar diquark computed using a symmetry-preserving regularisation of a vector$\times$vector contact interaction.
}
\end{figure}

The DAs in Eqs.\,\eqref{piDAans}, \eqref{qqDAans} are compared in Fig.\,\ref{FigCIDAs}.  Despite the similarities between their Bethe-Salpeter amplitudes, especially the fact that both are independent of the relative momentum between the valence constituents, the $0^+_{[ud]}$ DA is appreciably narrower than that of the pion.  The conspicuous difference between the $0^+_{[ud]}$ and $\pi$ is the mass obtained as a solution of their respective Bethe-Salpeter equations: $m_{0^+_{[ud]}} > m_\pi$.  Notably, contact interaction calculations of $0^+_{[ud]}$ and $\pi$ electromagnetic radii produce $r_{0^+} \approx 1.1 r_\pi$ \cite{Roberts:2011wy}.

\smallskip

\noindent\textbf{3.$\;$Rainbow-ladder kernel}.\,---\,%
In continuum studies of the had\-ron bound-state problem, the RL truncation has long been used to deliver predictions for observables \cite{Eichmann:2016yit, Qin:2020rad}.  With a judicious choice for the kernel, this symmetry-preserving truncation provides a good description of those systems wherein (\emph{i}) orbital angular momentum does not play a material role and (\emph{ii}) the non-Abelian anomaly can be neglected.  In all such cases, corrections to RL truncation interfere destructively, \emph{i.e}.\ largely cancel amongst themselves.

Regarding systems constituted from two valence bodies, one considers the Bethe-Salpeter equation:
\begin{equation}
[\Gamma(k;P)]_{tu} = \int_{dq}^\Lambda [\chi(q;P)]_{sr} {\mathpzc K(q,k;P)}^{rs}_{tu}, \label{EqBSE}
\end{equation}
where
$r\ldots u$ denote colour, flavour and spinor indices, and $\chi$ is defined through Eq.\,\eqref{defchi}.  For mesons, the RL truncation is specified via ($l = k_\eta -q_\eta= k_{\bar\eta}  - q_{\bar\eta}$)
\begin{subequations}
\label{EqRLInteraction}
\begin{align}
\label{KDinteraction}
\mathscr{K}_{tu}^{rs} & = {\mathpzc I}_{\mu\nu}(l) [i\gamma_\mu\frac{\lambda^{a}}{2} ]_{ts} [i\gamma_\nu\frac{\lambda^{a}}{2} ]_{ru}\,,\\
 {\mathpzc I}_{\mu\nu}(l)  & = \tilde{\mathpzc I}(l^2) T_{\mu\nu}(l)\,,
\end{align}
\end{subequations}
where $\{\lambda^a|a=1,\ldots,8\}$ are the generators of SU$(3)$-colour in the fundamental representation and $l^2 T_{\mu\nu}(l) = l^2 \delta_{\mu\nu} - l_\mu l_\nu$.

Expressing the diquark Bethe-Salpeter equation in terms of $\Gamma_{qq}(k_{\eta\bar\eta},P)C^\dagger$, the only change to Eq.\,\eqref{KDinteraction} is replacement of the second Gell-Mann matrix by its negative transpose \cite{Cahill:1987qr}: $\lambda^{a} \to [-\lambda^{a}]^{\rm T}$.  Now the origin of the one-half multiplicative factor mentioned above becomes clear, \emph{viz}.\
\begin{subequations}
\begin{align}
1_c \; \mbox{\rm meson:} & \quad \frac{\lambda^{a}}{2} {\mathbf I}_c \frac{\lambda^{a}}{2}  = \frac{4}{3}{\mathbf I}_c\,,\\
\bar 3_c \;\mbox{diquark:} &  \quad  \frac{\lambda^{a}}{2} i\lambda^{\mathpzc c}
\frac{[-\lambda^a]^{\rm T}}{2}  =  \frac{2}{3} i\lambda^{\mathpzc c} \,,
{\mathpzc c}=2,5,7\,.
\end{align}
\end{subequations}

At this point it is worth emphasising that diquark correlations are coloured.  Only in connection with partnering coloured objects (\emph{e.g}.\ quark or another diquark) is a colour singlet system obtained.  Consequently, diquarks are confined.  That is not true if RL truncation is used alone to develop the quark+quark scattering problem \cite{Maris:2002yu}.  Corrections to this leading-order truncation have been studied using an infrared-dominant interaction \cite{Munczek:1983dx}.  In fully self-consistent symmetry-preser\-ving analyses, such corrections purge bound-state poles from the quark+quark scattering matrix, whilst preserving the strong correlations \cite{Bender:1996bb, Bhagwat:2004hn}.  These studies signalled that, as coloured systems, like quarks and gluons, diquark propagation is effected by a compound two-point function whose analytic structure is unlike that of an asymptotic state \cite{Gribov:1999ui, Krein:1990sf, Bashir:2012fs, Brodsky:2012ku}.  Nevertheless, the propagation is characterised by a mass-scale commensurate with that obtained in RL analyses.

Predictions for meson and diquark properties follow immediately from numerical calculations once $\tilde{\mathpzc I}(l^2)$ is specified.  Experience has produced the following form \cite{Qin:2011xq} ($s=l^2$):
\begin{align}
\label{defcalG}
 \tfrac{1}{Z_2^2}\tilde{\mathpzc I}(s) & =
 \frac{8\pi^2}{\omega^4} D e^{-s/\omega^2} + \frac{8\pi^2 \gamma_m \mathcal{F}(s)}{\ln\big[ \tau+(1+s/\Lambda_{\rm QCD}^2)^2 \big]}\,,
\end{align}
where: $\gamma_m=12/(33-2N_f)$, $N_f=4$; $\Lambda_{\rm QCD}=0.234\,$GeV; $\tau={\rm e}^2-1$; and $s{\cal F}(s) = \{1 - \exp(-s/[4 m_t^2])\}$, $m_t=0.5\,$GeV.  Eq.\,\eqref{defcalG} preserves the one-loop renormalisation group behaviour of QCD.  Moreover, $0 < \tilde{\mathpzc I}(0) < \infty$, reflecting the fact that a nonzero gluon mass-scale appears as a consequence of EHM in QCD \cite{Boucaud:2011ug, Aguilar:2015bud, Huber:2018ned, Cui:2019dwv}.

The only parameters in Eq.\,\eqref{defcalG} are $D$, $\omega$.
For ground-state pseudo-scalar- and vector-mesons, $\omega=0.44\,$GeV, $D\omega$ $=$ $\varsigma^3$ $=$ $(0.8\,{\rm GeV})^3$ yield results for a raft of static and dynamic properties that are in good agreement with experiment, see \emph{e.g}.\ Refs.\,\cite{Gao:2016jka, Gao:2017mmp, Mojica:2017tvh, Chen:2018rwz, Ding:2018xwy, Xu:2019ilh, Ding:2019lwe, Cui:2020dlm, Cui:2020tdf}.  Notably, with $D\omega$ fixed, observables in these channels are practically unchanged on $\omega/{\rm GeV} \in [0.4,0.6]$ \cite{Qin:2011xq}.  Hence, in practice, Eq.\,\eqref{defcalG} is a one-parameter \emph{Ansatz}.
%


The kernel defined by Eqs.\,\eqref{EqRLInteraction}, \eqref{defcalG} preserves \linebreak QCD's ultraviolet behaviour; so, a renormalisation procedure must be implemented when solving the bound-state equations.  We use a mass-independent momentum subtraction scheme with renormalisation scale $\zeta=\zeta_H=0.33\,$GeV \cite{Cui:2020dlm, Cui:2020tdf}.  At $\zeta_H$, the dressed quasiparticles obtained from the quark gap equation express all properties of the bound state under consideration.  For instance, they carry all the hadron's momentum.   This approach ensures that parton splitting is properly expressed via $\zeta$-evolution of hadron wave functions \cite{Lepage:1979zb, Efremov:1979qk, Lepage:1980fj}, thereby eliminating a known deficiency of truncated bound-state kernels \cite{Ding:2019lwe, Cui:2020dlm, Cui:2020tdf}.

\smallskip

\noindent\textbf{4.$\;$DAs using a realistic interaction}.\,---\,%
%
The asymptotic profile for the leading-twist two-parton DA in QCD is \cite{Lepage:1979zb, Efremov:1979qk, Lepage:1980fj}:
\begin{equation}
\varphi_{\rm as}(x) = 6 x \bar x\,.
\end{equation}
The first two independent Mellin moments of this distribution are:
\begin{equation}
\label{asM24}
\begin{array}{c|c|c}
       & \langle x^2 \rangle & \langle x^4 \rangle  \\\hline
\varphi_{\rm as}\ & \ \tfrac{3}{10} = 0.3\ & \ \tfrac{1}{7} = 0.143
\end{array}\,.
\end{equation}

\smallskip

\noindent\emph{4.1$\;$Pion and scalar diquark}.\,---\,%
Using the kernel described in Sect.\,3 and working in the isospin-symmetry limit, we solved the quark gap and pion Bethe-Salpeter equations.  Following Ref.\,\cite{Maris:1997tm}, the numerical procedures have evolved to the point where this exercise is straightforward.  The algorithm improvements in Ref.\,\cite{Krassnigg:2009gd} were a valuable step along the way.

With a renormalisation point invariant current-quark mass $\hat m = 7.2\,$MeV, which corresponds to a one-loop evolved current-mass $m^{\zeta_2} = 5.0\,$MeV at $\zeta_2=2\,{\rm GeV}$, one obtains the pion mass and decay constant in Table~\ref{TabStatic}.  The decay constant is obtained by computing the zeroth Mellin moment of both sides in Eq.\,\eqref{varphiresult}.

\begin{table}[t]
\caption{\label{TabStatic}
Static properties of mesons and diquarks evaluated with the bound-state kernels described in Sect.\,3. The current mass $m^{\zeta_2} = 5.0\,$MeV, which is commensurate with other estimates \cite{Zyla:2020zbs}.
For comparison, meson empirical values are \cite{Zyla:2020zbs} (in GeV): $m_\pi = 0.138$, $f_\pi = 0.0924$, $m_\rho=0.775$, $f_\rho = 0.153$.
(\emph{N.B}.\ $\pi$ and $0^+_{[ud]}$ do not possess a transverse decay constant.
``$[\cdot]$'' entries in decay constant rows list $f_{\rm meson}/\surd 3$ or $f_{\rm diquark}/\surd 2$.  All dimensioned quantities in GeV.)}
\begin{center}
\begin{tabular*}
{\hsize}
{
l@{\extracolsep{0ptplus1fil}}
|l@{\extracolsep{0ptplus1fil}}
l@{\extracolsep{0ptplus1fil}}
l@{\extracolsep{0ptplus1fil}}
l@{\extracolsep{0ptplus1fil}}}\hline
 ch\        & $\pi\ $ & $0^+_{[ud]}\ $ & $\rho\ $ & $1^+_{\{u d\}}\ $ \\\hline
$m_{\rm ch}\ $ & $0.14\phantom{3}\ $ & $0.89\phantom{3}\ $ & $0.72\ $& $1.04\ $\\
$f_{\rm ch}\ $ & $0.091$\,$[0.053]\ $ & $0.072$\,$[0.051]\ $ & $0.14$\,$[0.083]\ $ & $0.088$\,$[0.062]\ $   \\
$f_{\rm ch}^\perp\ $ & & & $0.11$\,$[0.063]\ $ & $0.054$\,$[0.038]\ $ \\\hline
\end{tabular*}
\end{center}
\end{table}

Solving the analogous RL $0^+_{[ud]}$ Bethe-Salpeter equation yields the associated mass and decay constant in Table~\ref{TabStatic}.  As usual, the ``decay constant'', obtained by computing the zeroth Mellin moment of both sides in Eq.\,\eqref{varphiud}, has the meaning of a value for the pseudovector projection of the $0^+_{[ud]}$ wave function onto the origin in configuration space.  Owing to their different colour structures, the natural comparison is $ f_\pi / \surd 3 \leftrightarrow f_{[ud]}/\surd 2$.  These values are listed in square brackets in the table.

With numerical solutions for the quark propagators and meson/diquark Bethe-Salpeter amplitudes in hand, one can use Eqs.\,\eqref{varphiresult} or Eqs.\,\eqref{varphiud} to compute DA Mellin moments of nonzero order.  Their direct calculation is possible using the numerical technique described in Ref.\,\cite{Ding:2015rkn}.  For the light-quark systems considered herein, the scheme is adequate for $n\leq 4$.  Since all systems considered have DAs that are symmetric around $x=1/2$, precision of the numerical procedure can be ensured by checking the following identities:
\begin{equation}
\langle x \rangle_{\rm ch} = \tfrac{1}{2}\,,\;
\langle x^3 \rangle_{\rm ch} = \tfrac{3}{2} \langle x^2 \rangle_{\rm ch} - \tfrac{1}{4}\,.
\end{equation}

\begin{figure*}[!t]
\hspace*{-1ex}\begin{tabular}{ll}
{\sf A} & {\sf B} \\[0.1ex]
\includegraphics[clip, width=0.48\textwidth]{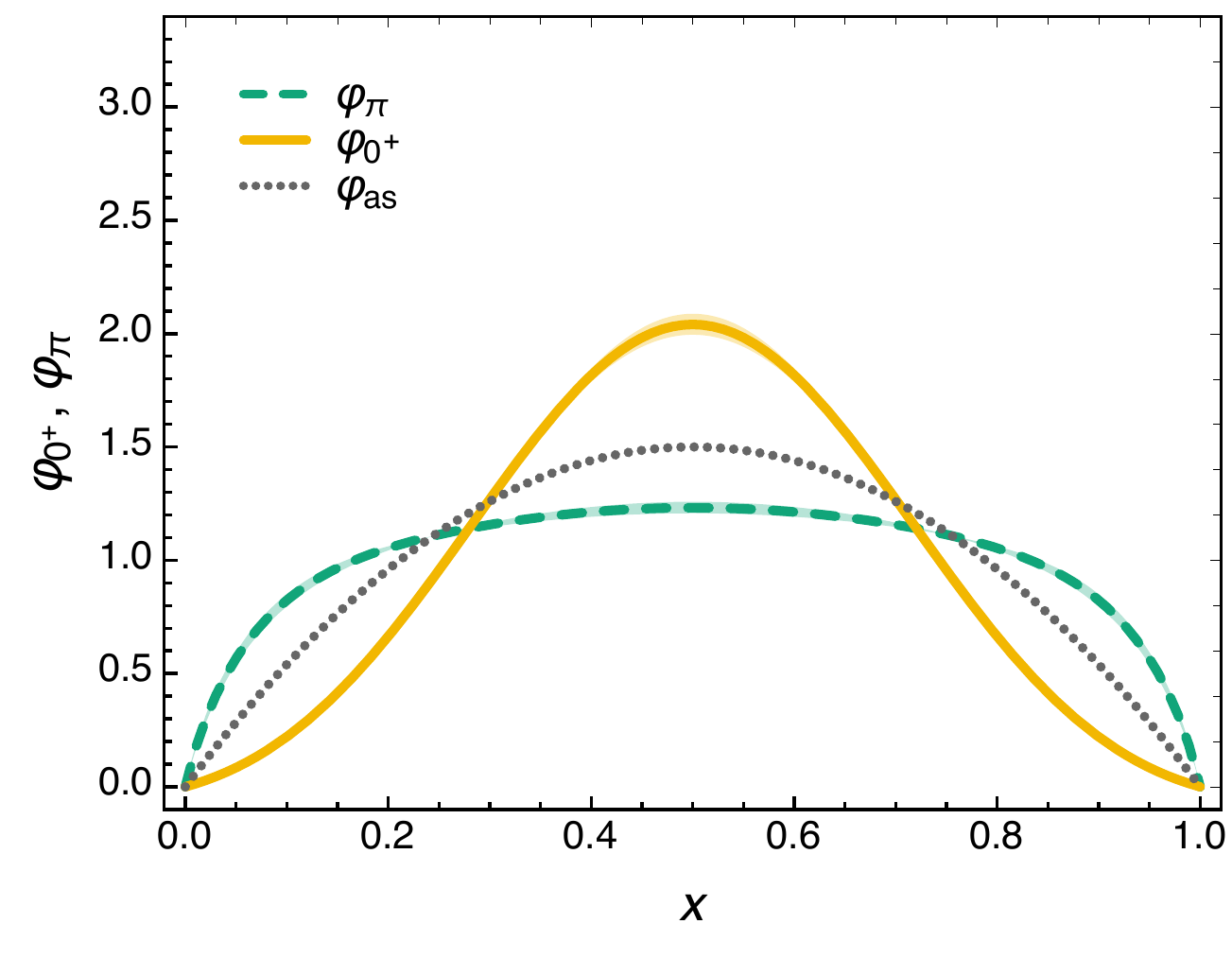} &
\includegraphics[clip, width=0.48\textwidth]{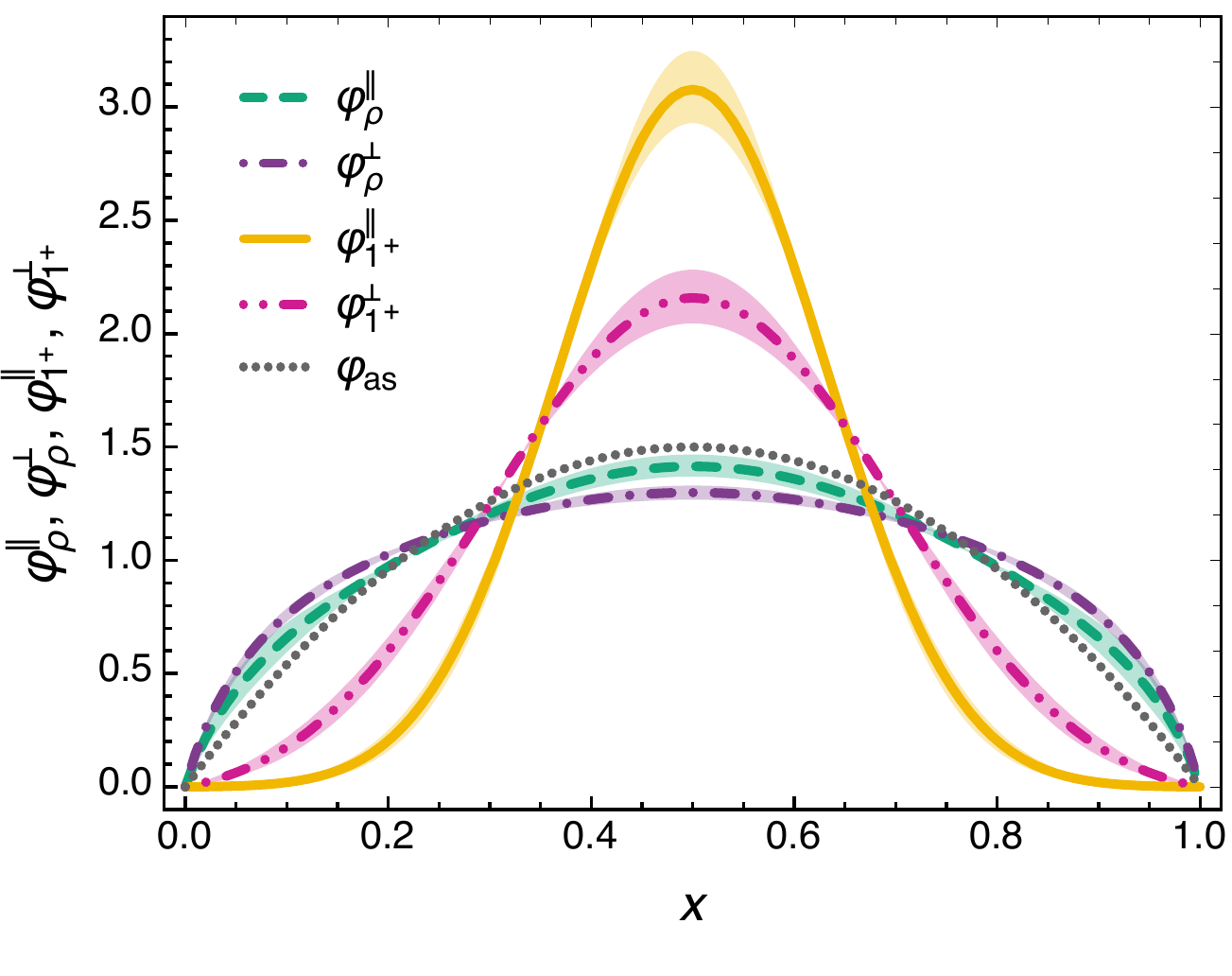}
\end{tabular}
\caption{\label{DrawDApi}
\emph{Left panel}\,--\,{\sf A}.  Two-particle DAs for the pion and scalar diquark computed using the QCD-connected RL kernel described in Sect.\,3.
\emph{Right panel}\,--\,{\sf B}.  Similarly for the two-particle DAs of the $\rho$-meson and pseudovector diquark.
In both panels, the asymptotic profile ``as'' is included for reference.
}
\end{figure*}

Access to moments $n>4$ is possible using the perturbation theory integral representation scheme introduced for this purpose elsewhere \cite{Chang:2013pq}.  However, experience has shown that when combined with physical requirements, such as the known $x\simeq 0, 1$ endpoint behaviour and unimodality for the DAs of ground-state pseudoscalar and vector mesons, the first four moments are sufficient to reconstruct a sound pointwise DA approximation.  (Unimodality is violated by scalar and some pseudovector meson ground-state DAs and by the DAs of radially-excited mesons \cite{Li:2016mah, Li:2016dzv}.)

Results for the first two nontrivial $\pi$ and $0^+_{[ud]}$ DA moments are listed here:
\begin{equation}
\label{DAMoments}
\begin{array}{l|ll}
     & \ \pi & \ [ud] \\\hline
\langle x^2 \rangle\  & \ 0.315 & 0.284 \\
\langle x^4 \rangle\  & \ 0.159(9) & 0.115(4) \\
\end{array}\,.
\end{equation}
Beyond quadrature error, which is negligible, there is no uncertainty in $\langle x^2\rangle$.  The uncertainty in $\langle x^4\rangle$ arises from extrapolation, following the algorithm in Ref.\,\cite{Ding:2015rkn}.

Comparisons with Eq.\,\eqref{asM24} suggest immediately that $\varphi_\pi$ is broader and flatter than $\varphi_{\rm as}$ because the moments of $\varphi_\pi$ are greater than those of $\varphi_{\rm as}$ and larger moments indicate more support on the endpoint domains.  Conversely, $\varphi_{0^+}$ must be narrower and taller than $\varphi_{\rm as}$.

With these indications in mind, we chose the following reconstruction functions \cite{Ding:2015rkn, Ding:2019lwe}:
\begin{subequations}
\label{DAformspi}
\begin{align}
\varphi_\pi(x) & = {\mathpzc n}_\pi x \bar x \left[1 + \alpha_\pi \sqrt{x\bar x} + \beta_\pi x\bar x\right]\,,\\
\varphi_{0^+}(x) &=  {\mathpzc n}_{0^+} x \bar x \exp\left[-a^2_{0^+} (x-\bar x)^2 \right]\,,
\end{align}
\end{subequations}
where the computed values of $ {\mathpzc n}_{\pi,0^+}$ ensure unit normalisation, Eqs.\,\eqref{unitnormpi}, \eqref{unitnormud}.  The moments in Eq.\,\eqref{DAMoments} are reproduced using:
\begin{equation}
\label{DAformspi2}
\begin{array}{l|c|c|c}
& \alpha_\pi & \beta_\pi & a_{0^+}\\\hline
{\rm up} & -2.52  & \ 2.00 \ & 1.43\\
{\rm mid}\ & -2.57  & \ 2.03 \ & 1.37\\
{\rm low} & -2.64  & \ 2.10 \ & 1.31
\end{array}\,,
\end{equation}
where the ``up'' values yield the smallest result for $\langle x^4\rangle$ in Eq.\,\eqref{DAMoments}, etc.; hence, the tallest DA at $x=\tfrac{1}{2}$.

The $\pi$ and $0^+_{[ud]}$ DAs obtained using Eqs.\,\eqref{DAformspi}, \eqref{DAformspi2} are depicted in Fig.\,\ref{DrawDApi}A.  As anticipated and reproducing the result in Ref.\,\cite{Chang:2013pq}, the pion's DA is a broad, concave function, which is both narrower and flatter than $\varphi_{\rm as}(x)$.  On the other hand, the $0^+_{[ud]}$ DA is narrower than $\varphi_{\rm as}(x)$; so, as in the contact interaction study, much narrower and taller than $\varphi_\pi(x)$.  Here, too, computations of the electromagnetic radii of these systems produce $r_{0^+} \approx 1.1 r_\pi$ \cite{Maris:2004bp}.

\smallskip

\noindent\emph{4.2$\;$ $\rho$-meson and pseudovector diquark}.\,---\,%
The above analysis can be repeated for $J=1$ systems following Refs.\,\cite{Ding:2015rkn, Gao:2014bca}, yielding the masses and decay constants in Table~\ref{TabStatic}.  (Since we assume isospin symmetry, it is unnecessary to distinguish between $\{uu\}$ and $\{ud\}$ diquarks.)  In this case, there are two decay constants and an associated DA for each state considered: $\varphi^{\parallel, \perp}$, describing, respectively, the light-front fraction of the system's total momentum carried by a quark in either a light-front longitudinally or transversely polarised state.  For the $\rho$-meson, they are obtained via
\begin{subequations}\label{vectormesonDAs}
\begin{align}
n\cdot P \, f_\rho & \varphi_\rho^\parallel(x;\zeta) =
m_\rho N_c {\rm tr}_{\rm D}Z_2(\zeta,\Lambda) \nonumber\\
& \quad \times \int_{dk}^\Lambda \delta_n^x(k_\eta)\,
\gamma\cdot n n_\nu \chi_\nu^\rho  (k_{\eta\bar\eta},P ;\zeta) \,, \\
f^\perp_\rho & \varphi^\perp_\rho(x;\zeta)  =
- \tfrac{1}{2} N_c {\rm tr}_{\rm D}Z_T(\zeta,\Lambda) \nonumber\\
& \quad \times \int_{dk}^\Lambda \delta_n^x(k_\eta)\,
n_\mu \sigma_{\mu\alpha} O^\perp_{\alpha\nu}
\chi_\nu^\rho  (k_{\eta\bar\eta},P ;\zeta) \,,
\end{align}
\end{subequations}
where $ O^\perp_{\alpha\nu} = \delta_{\alpha\nu} + n_\alpha \bar n_\nu + \bar n_\alpha n_\nu$ and
$Z_T(\zeta,\Lambda)$ is the quark tensor-vertex renormalisation constant \cite[Appendix\,A]{Gao:2014bca}.
%
%
The expressions for the $1^+_{\{ud\}}$ DAs are obvious by analogy, remembering that the colour factor is ``2'' in this case.

The lowest independent Mellin moments of the $\rho$ and $1^+_{\{ud\}}$ DAs are:
\begin{equation}
\label{DAMomentsJ1}
\begin{array}{c|cccc}
     & \rho_\parallel & \rho_\perp & \{ud\}_\parallel & \{ud\}_\perp \\\hline
\langle x^2 \rangle\  & \ 0.306(3) & 0.310(1) & 0.266(2) & 0.281(2)\\
\langle x^4 \rangle\  & \ 0.155(9) & 0.158(6) & 0.108(8) & 0.110(8)\\
\end{array}\,.
\end{equation}
The complexity of the Bethe-Salpeter amplitude for $J=1$ systems, highlighted elsewhere \cite[Sec.\,V.B]{Maris:1999nt}, means that extrapolation is also required to obtain $\langle x^2\rangle$ in these cases; hence, an uncertainty is listed.

Comparisons between the values in Eq.\,\eqref{DAMomentsJ1} and those in Eqs.\,\eqref{asM24}, \eqref{DAMoments} suggest the following ordering of DAs:
\begin{subequations}
\label{rhoordering}
\begin{align}
\varphi_\pi >_B \ &                                
\varphi_\rho^\perp >_B  \                 
\varphi_\rho^\parallel >_B  \           
\varphi_{\rm as}    \\                 
\varphi_{\rm as} >_B \ &
\varphi_{[ud]} \gtrsim_B \             
\varphi_{\{ud\}}^\perp   >_B \            
\varphi_{\{ud\}}^\parallel          
\end{align}
\end{subequations}
where $>_B$ indicates broader-and-flatter-than.  Using \linebreak Eqs.\,\eqref{DAformspi}, the moments in Eq.\,\eqref{DAMomentsJ1} are reproduced by the following coefficients:
\begin{equation}
\label{DAformspi3}
\begin{array}{l|c|c|c|c|c|c}
& \alpha_\rho^\parallel & \beta_\rho^\parallel &
    \alpha_\rho^\perp & \beta_\rho^\perp &  a_{1^+}^\parallel &  a_{1^+}^\perp \\\hline
{\rm up} & -1.99\  & \ 2.07 \ & -2.45\ & \ 2.08\ & \ 2.67\ & \ 1.67\ \\
{\rm mid}\ & -2.57\  & \ 2.56 \ & -2.49\ & \ 2.05\ & \ 2.51\ & \ 1.52\ \\
{\rm low} & -2.99\  & \ 3.00 \ & -2.52\ & \ 2.02\ & \ 2.36\ & \ 1.38\
\end{array}\,.
\end{equation}

The $\rho$ and $1^+_{\{ud\}}$ DAs obtained using Eqs.\,\eqref{DAformspi}, \eqref{DAformspi3} are drawn in Fig.\,\ref{DrawDApi}B.  The ordering of peak heights matches that anticipated in Eq.\,\eqref{rhoordering}, with the $\rho$-meson results being consistent with those in Ref.\,\cite{Gao:2014bca}.
Continuing the pattern disclosed in Sects.\,2, 3, the diquark DA is also narrower and taller than that of its counterpart meson for $J=1$ systems.

\smallskip

\noindent\textbf{5.$\;$Summary and perspective}.\,---\,%
Using the leading-order (rainbow-ladder, RL) truncation of those equations needed to solve continuum two-valence-body bound-state problems, we computed the leading-twist two-dres\-sed-parton distribution amplitudes (DAs) of light-quark (\emph{a}) pseudoscalar mesons and scalar diquarks and (\emph{b}) vector mesons and pseudovector diquarks.  In both cases, the diquark DAs are narrower and taller than those of their meson counterparts; moreover, all diquark DAs are narrower and taller than the QCD asymptotic meson profile $\varphi_{\rm as}(x) = 6 x(1-x)$.  Thus, the valence quasiparticles in a diquark are less likely to carry a large light-front fraction of the system's total momentum than those in a meson.  One can expect to see manifestations of these characteristics in baryon DAs \cite{Mezrag:2017znp}.

Given the relation between valence quasiparticle DAs and distribution functions (DFs) \cite{Roberts:2021nhw}:
\begin{equation}
\label{PDFeqPDA2}
{\mathpzc q}(x;\zeta_H) \approx \varphi^2(x;\zeta_H)\,,
\end{equation}
then the features of diquark DAs revealed herein will also be transferred into their DFs.  Hence, diquark DFs are likely narrower than those of their counterpart mesons.  (Explicit checks of this expectation are underway.) That being so, then valence quarks sequestered within a diquark correlation inside a proton will be less likely to participate in a hard interaction than the bystander valence quark.  Such effects should be manifest in the ratio of $d$ and $u$ quark DFs in the proton.  Our analysis may therefore be useful in informing future calculations of baryon DFs.

As noted above, we employed RL truncation in this study, which is state-of-the-art for \emph{ab initio} studies of baryon properties in the continuum \cite{Barabanov:2020jvn, Eichmann:2016yit, Qin:2020rad}.  Notwithstanding that, owing to confinement, certain aspects of diquark correlations may be affected by corrections to this leading order truncation.  Accordingly, it is worth replicating the analysis described herein using the more sophisticated bound-state kernels that have recently been developed \cite{Williams:2015cvx, Binosi:2016rxz, Qin:2020jig}.

\smallskip

%
\noindent\textbf{Acknowledgments}.\,---\,%
We are grateful for constructive comments from
Z.-F.~Cui, Z.-N.~Xu, P.-L.~Yin and J.-L.~Zhang.
Work supported by:
Jiangsu Pro\-vince \emph{Hundred Talents Plan for Professionals};
and
National Natural Science Foundation of China (grant 11822503).
%




\end{document}